\documentclass[11pt]{article}

\usepackage{amssymb,epsfig,fullpage}

\def\arctanh{\mathop{\textrm{arctanh}}\nolimits}

\begin{document}

\title{Stability and bifurcations in a model of antigenic variation in malaria
\thanks{This work was partially supported by the ATRJVVO
grant from the James Martin 21st Century School, University of
Oxford.}
}

\author{K.B. Blyuss\thanks{Corresponding author. Email: k.blyuss@sussex.ac.uk}
\\\\ Department of Mathematics, University of Sussex,\\Brighton, BN1 9QH, United Kingdom
\and Sunetra Gupta
\\\\ Department of Zoology, University of Oxford,\\ Oxford, OX1 3PS, United Kingdom
}

\maketitle

\begin{abstract}
We examine the properties of a recently proposed model for antigenic
variation in malaria which incorporates multiple epitopes and both
long-lasting and transient immune responses. We show that in the
case of a vanishing decay rate for the long-lasting immune response,
the system exhibits the so-called ``bifurcations without
parameters'' due to the existence of a hypersurface of equilibria in
the phase space. When the decay rate of the long-lasting immune
response is different from zero, the hypersurface of equilibria
degenerates, and a multitude of other steady states are born, many
of which are related by a permutation symmetry of the system. The
robustness of the fully symmetric state of the system was
investigated by means of numerical computation of transverse
Lyapunov exponents. The results of this exercise indicate that for a
vanishing decay of long-lasting immune response, the fully symmetric
state is not robust in the substantial part of the parameter space,
and instead all variants develop their own temporal dynamics
contributing to the overall time evolution. At the same time, if the
decay rate of the long-lasting immune response is increased, the
fully symmetric state can become robust provided the growth rate of
the long-lasting immune response is rapid.
\end{abstract}

\section{Introduction}

Several pathogens, including {\it Plasmodium
falciparum} malaria and African trypanosomes, achieve immune
escape by the so-called {\it antigenic variation} (see a recent review
by Gupta \cite{Gu}). The latter essentially refers to a process by which a pathogen keeps
changing its surface proteins, thus preventing antibodies from
recognizing and destroying it. Antigenic
variation is achieved by exploiting a large repertoire of antigenic
variants that differ in some of their epitopes.
An important requirement here is that the variants must not be expressed
all at the same time, as otherwise the resulting immune response
will detect and destroy all of them, thus terminating the infection.

Here we examine a particular model of antigenic variation for {\it P.
falciparum} malaria, put forward by Recker {\it et al.} \cite{RNBKMNG}.
Within this framework, each variant is assigned one major epitope, which is
unique to that variant, and also several minor epitopes that are
shared between different variants.  Both types of epitopes elicit epitope-specific responses,
but in the case of the minor epitopes these are cross-protective between variants that share
them. A critical feature of the model is that the immune response to the major epitope
(uniquely variant-specific) is long-lasting in comparison with the immune responses
(frequently cross-protective) to the minor epitopes. Under these conditions, the dynamics may
be characterized by sequential domination of different variants. Thus,
the conclusion of the model is that effectively
the host immune system can itself be responsible for prolonging
the malaria infection and causing chronicity. Through numerical simulations and by analysing
a caricature of the model involving complete synchrony between
variants, Recker and Gupta \cite{RG} have shown that stronger cross-protective immune
responses lead to prolonged length of infection and reduced severity of the disease.
This was explained by
the conflicting interaction of cross-protective and variant-specific immune responses.

In this paper we perform a detailed study of this model with
particular emphasis on stability aspects, as well as possible
bifurcation scenarios. First, we consider the case when the
variant-specific immune responses to the major epitopes do not decay. In
this case, the phase space of the system possesses a very peculiar
geometry with a high-dimensional surface of equilibria having
different types of stability. We show, it is exactly this
curious structure that causes successive re-appearance of
different malaria variants in the dynamics until the specific
immune responses reach sufficient protective level to prevent
further appearance of given variants in the dynamics. If the
specific immune responses can decay (even slightly), the dynamics
is qualitatively different, as the phase space geometry changes
significantly. Now it contains a large number of distinct
equilibria with different number of non-zero variants, some
of which can be related by the permutation symmetry of the system.

An interesting tool for investigating the dynamics is the imposition of
synchrony among the variants. From mathematical perspective, in the
case of complete synchrony the dimension of the system is
drastically reduced. Here, we  use the tools of synchronization
theory to investigate the robustness of such state.

The outline of this paper is as follows. In the next section the
model of cross-reactive immune response to malaria is introduced
and its basic properties are discussed. Section 3 contains the
analysis of a particular case when the decay rate of a
long-lasting immune response vanishes. Numerical simulations will
be presented that illustrate the behaviour of the system in this
case. A general situation of arbitrary non-decaying specific
immune responses is considered in Section 4. In section 5 the
stability of the fully symmetric state of the system is
investigated by means of numerical computation of transverse
Lyapunov exponents. The paper concludes in section 6 with a
discussion.

\section{Model definition}

In this section we use the above-mentioned multiple epitope
description to introduce a model of the interaction of malaria
variants with the host immune system. Our derivation follows that
of Recker {\it et al.} \cite{RNBKMNG} with some refinements.

It is assumed that each antigenic variant $i$ consists of a single
unique major epitope, that elicits a long-lived (specific) immune
response, and also of several minor epitopes that are not unique
to the variant. Assuming that all variants have the same net
growth rate $\phi$, their temporal dynamics is described by the
equation
\begin{equation}\label{yeq}
\frac{dy_i}{dt}=y_i(\phi -\alpha z_i-\alpha' w_i),
\end{equation}
where $\alpha$ and $\alpha'$ denote the rates of variant
destruction by the long-lasting immune response $z_i$ and by the
transient immune response $w_i$, respectively, and index $i$ spans
all possible variants. The dynamics of the variant-specific immune
response can be written in its simplest form as
\begin{equation}\label{zeq}
\frac{dz_i}{dt}=\beta y_{i}-\mu z_{i},
\end{equation}
with $\beta$ being the proliferation rate and $\mu$ being the
decay rate of the immune response. Finally, the transient
(cross-reactive) immune response can be described by the minor
modification of the above equation (\ref{zeq}):
\begin{equation}\label{weq}
\frac{dw_i}{dt}=\beta'\sum_{j\sim i} y_j-\mu' w_i,
\end{equation}
where the sum is taken over all variants sharing the epitopes with
the variant $y_i$. We shall use the terms long-lasting and
specific immune response interchangeably, likewise for transient
and cross-reactive.

To formalize the above construction, one can introduce the
adjacency matrix $A$, whose entries $A_{ij}$ are equal to one if
the variants $i$ and $j$ share some of their minor epitopes and
equal to zero otherwise. Obviously, the matrix $A$ is always a
symmetric matrix. Prior to constructing this matrix it is
important to introduce a certain ordering of the variants
according to their epitopes. For this purpose we shall use the
{\it lexicographic} ordering, as explained below. To illustrate
this, suppose we have a system of two minor epitopes with two
variants in each epitope, which is the simplest non-trivial system
of epitope variants. In this case, the total number of variants is
four, and they are enumerated as
\begin{equation}\label{var4}
\begin{array}{l}
1\hspace{1cm}11\\
2\hspace{1cm}12\\
3\hspace{1cm}21\\
4\hspace{1cm}22\\
\end{array}
\end{equation}
It is clear that for a system of $m$ minor epitopes with $n_{i}$
variants in each epitope, the total number of variants is given by
\begin{equation}
N=\prod_{i=1}^{m}n_{i}.
\end{equation}

Now that the ordering of variants has been fixed, it is an easy
exercise to construct the adjacency matrix $A$ of variant
interactions. For the particular system of variants (\ref{var4}),
this matrix has the form
\begin{equation}\label{mat4}
A=\left(
\begin{array}{cccc}
1&1&1&0\\
1&1&0&1\\
1&0&1&1\\
0&1&1&1
\end{array}
\right).
\end{equation}
In general, for a system of two minor epitopes with $m$ variants
in the first epitope and $n$ variants in the second, the matrix
$A$ will be an $mn\times mn$ block matrix consisting of $n\times
n$ blocks of ones along the main diagonal, with the rest of the
matrix being filled with $n\times n$ identity matrices. For
simplicity, in the rest of the paper we will concentrate on the
case of two minor epitopes, but the results can easily be
generalized for arbitrary number of minor epitopes. Using the
adjacency matrix one can rewrite the system
(\ref{yeq})-(\ref{weq}) in a vector form
\begin{equation}\label{vs}
\frac{d}{dt} \left(\begin{array}{l}{\bf y}\\
{\bf z}\\{\bf w}\end{array}\right) =F({\bf y},{\bf z},{\bf w})=
\left\{
\begin{array}{l}
{\bf y}(\phi{\bf 1}_{N}-\alpha{\bf z}-\alpha'{\bf w}),\\
\beta{\bf y}-\mu{\bf z},\\
\beta'A{\bf y}-\mu'{\bf w},
\end{array}
\right.
\end{equation}
where ${\bf y}=(y_1,y_2,...,y_{N})$ etc., ${\bf 1}_{N}$ denotes a
vector of the length $N$ with all components equal to one, and in
the right-hand side of the first equation multiplication is taken
to be entry-wise so that the output is a vector again.
\begin{figure}
\hspace{3cm}\epsfig{file=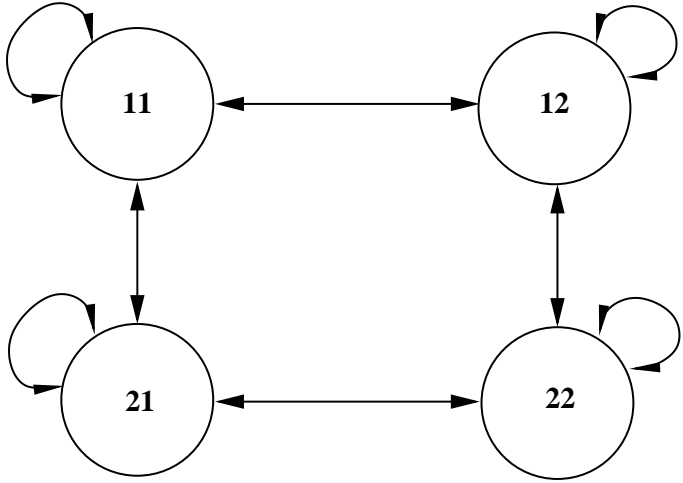,width=9cm}
\caption{Interaction of malaria variants in the case of two minor
epitopes with two variants in each epitope.}
\label{fig1}
\end{figure}

To better understand the symmetry of the system it is convenient
to represent graphically relations between different variants.
Figure~\ref{fig1} shows such relations in the case of two minor
epitopes. One can observe that within each horizontal and each
vertical stratum, the network of variants is characterized by an
``all-to-all" coupling \cite{DS}. Besides this, if the number of
variants in both minor epitopes is the same, then there is an
additional reflectional symmetry. Formally this means the system
is equivariant with respect to the following symmetry group
\cite{GS}
\begin{equation}
\mathcal{G}=\left\{
\begin{array}{l}
{\bf S}_{m}\times {\bf S}_{n},\mbox{ }m\neq n,\\
{\bf S}_{m}\times {\bf S}_{m}\times{\bf\mathbb{Z}}_{2},\mbox{
}m=n.
\end{array}
\right.
\end{equation}
This construction can be generalized in a straightforward way for
a larger number of minor epitopes. It is noteworthy that each of the variants has
exactly the same number of connections to other variants.

We finish this section by noting that the system (\ref{vs}) is well-posed, in that
provided the initial conditions for this system are non-negative ${\bf y}(0)\geq 0,{\bf z}(0)\geq 0,{\bf w}(0)\geq 0$, the solutions satisfy ${\bf y}(t)\geq 0,
{\bf z}(t)\geq 0, {\bf w}(t)\geq 0$, for all $t\geq 0$.\\

\noindent{\bf Remark.} In many cases it is reasonable to assume
the initial conditions for the system (\ref{vs}) to be of the form
$({\bf y}\geq 0,{\bf z}=0,{\bf w}=0)$. A possible exception is
when the immune system has already built-up a long-lasting
response from prior exposure to a certain variant. In this case,
the initial condition for the system (\ref{vs}) will contain
non-zero entries for some of ${\bf z}$ variables \cite{RG}.

\section{The case of non-decaying specific immune response}

We begin our analysis of the system (\ref{vs}) by considering a
particular case of vanishing decay rate of the long-lasting immune
response $\mu=0$ (some partial results for this case have been
obtained in \cite{RNBKMNG}). In this case the only steady states
of this system are given by
\begin{equation}
{\bf y}={\bf w}=0,\mbox{ }{\bf z}={\rm const}.
\end{equation}
This is a rather degenerate situation as the fixed points are not
separated in the phase space, but rather form an $N$-dimensional
hypersurface with each point of it being a fixed point of the
system (\ref{vs}). Linearization near one such fixed point has the
eigenvalues $(-\mu')$ of multiplicity $N$, zero of multiplicity
$N$, and the rest of the spectrum is given by
\begin{equation}
\phi-\alpha z_1,\phi-\alpha z_2,...,\phi-\alpha z_N.
\end{equation}
The generalized eigenvectors of the zero eigenvalue correspond to
the $N$ directions along the hypersurface of the fixed points. As
long as there is at least one $z_i<\phi/\alpha$, the corresponding
steady state is a saddle, otherwise it is a stable node. From the
dynamical systems perspective, the case $\mu=0$ corresponds to the
so-called {\it bifurcations without parameters} \cite{FLA}.
Indeed, in the space $\{{\bf y}={\bf w}=0\}$ as one crosses the
hyperplane $z_{j}=\phi/\alpha$, one of the eigenvalues crosses
zero along the real axis. Furthermore, since the hypersurface
$\{{\bf y}={\bf w}=0\}$ is in general high-dimensional, the cases
of two or more eigenvalues crossing zero at the same time (this
happens along the lines $z_i=z_j=\phi/\alpha,i\neq j$) are still
generic, and these lead to ``bifurcations'' of a higher
co-dimension. It is important to note that all these bifurcations
are of the steady state type and there is no possibility of a Hopf
bifurcation that could lead to temporally periodic solutions.

\begin{figure}
\hspace{-1cm}\epsfig{file=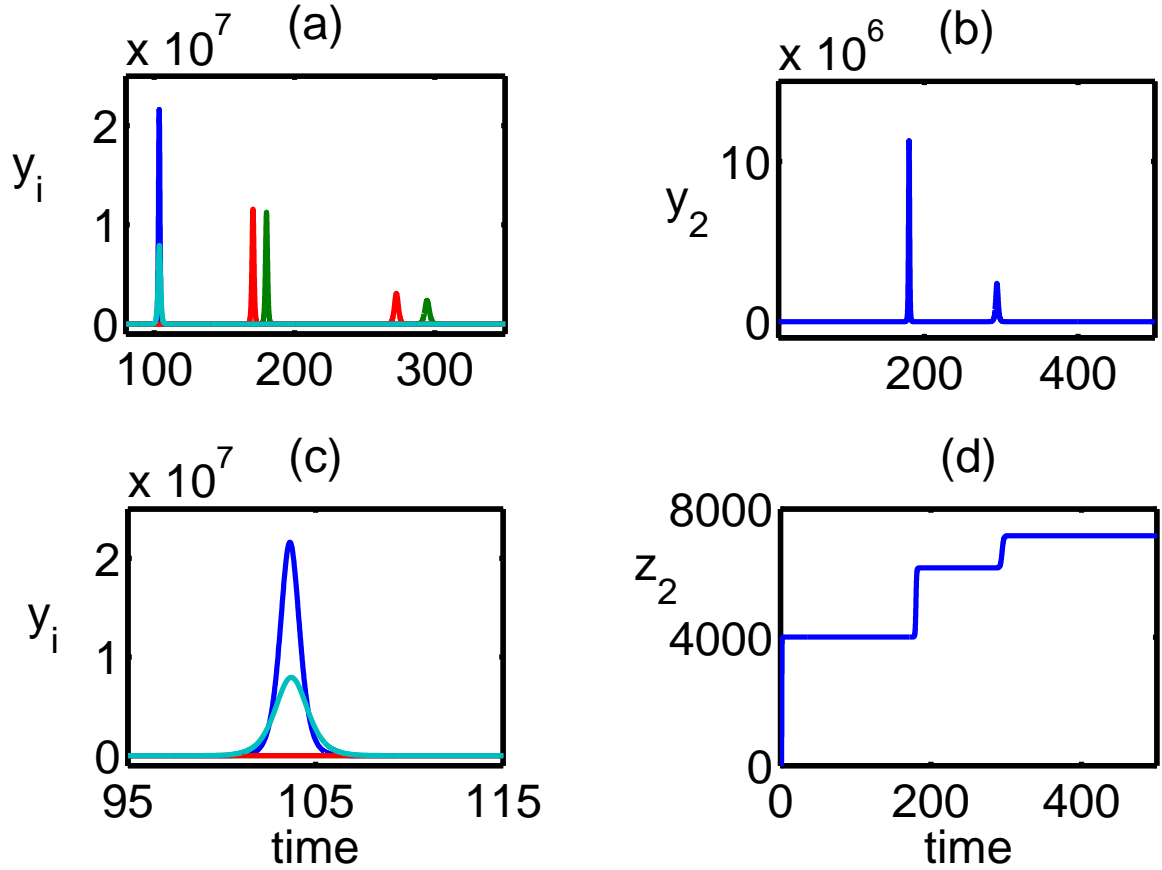,width=17cm}
\caption{Temporal
dynamics of the system (\ref{vs}) with two variants in each of the
two epitopes. Parameter values are
$\alpha=\alpha'=10^{-3},\beta=\beta'=10^{-4},\mu=0,\mu'=0.02,\phi=7.5$
\cite{RNBKMNG}. a) Time evolution of the variants $y_{i}$. b)
Dynamics of a single variant $y_{1}$. c) The close-up of the
initial stage of evolution of the variants. d) Dynamics of the
long-lasting immune response $z_{1}$.}\label{fig2}
\end{figure}

Figure~\ref{fig2} shows numerical simulations of a typical
behaviour in the system (\ref{vs}) for $\mu=0$. These results were
obtained by integrating the system (\ref{vs}) using the variable
order solver based on backward differentiation formulas to account
for the stiffness of the system \cite{SR}. Initially most variants have quite high amplitudes, but as the
time progresses, their amplitudes decrease as illustrated in Fig.~\ref{fig2}(a).
Figures (b) and (d) illustrate this feature in more detail by showing the dynamics of a single variant and its
specific immune response. With each subsequent re-appearance of the variant, the specific immune response to it is building up, and ultimately it reaches a protective level, which prevents this variant from 
ever re-appearing in the dynamics. As suggested by Fig.~\ref{fig2}(c), sometimes more than one variant appear at the same time, and this is very good from the immune system perspective, as it allows simultaneous destruction of all of these variants. The question of synchronization between different variants will be investigated in Section 5.

The fact that the system exhibits the jumps from one variant to
another can be explained by the existence of the above-mentioned
hypersurface of equilibria. When one of the variants decays, the
trajectory approaches the neighbourhood of the hypersurface of
equilibria, and since all points on this hypersurface are saddles
of different dimensions, the trajectory is pushed away along the
unstable manifold of one of these fixed points. This behaviour is
reminiscent of that in the neighbourhood of a heteroclinic cycle
\cite{AF,PD}, with the major difference being that in the present
case the nodes of the cycle are not distinct but rather form a
smooth hypersurface. There is a clear separation of time scales in
the dynamics: the trajectories move quickly to/away from the
invariant ${\bf w}-{\bf z}$ plane, and then they slowly move
towards the hyper-axis ${\bf y}={\bf w}=0$ before the next
iteration. With time, the phase space excursions between
subsequent returns to the equilibrium manifold become shorter
(they are restricted by the ever growing $z$ variables), and
eventually all trajectories converge to a point ${\rm\bf T}=({\bf y}={\bf
w}=0,{\bf z}=(\phi/\alpha){\bf 1 }_{N})$. Similar behaviour takes place in the phase coordinates of other variants, which all approach the point ${\rm\bf T}$. The importance of such a point for
understanding the dynamics has been previously highlighted, for
instance, in the analysis of adaptive control systems, where it
gave rise to bad point bifurcations at which the close-loop
systems could never be stabilised \cite{RTO}.

\begin{figure}
\hspace{-1cm}\epsfig{file=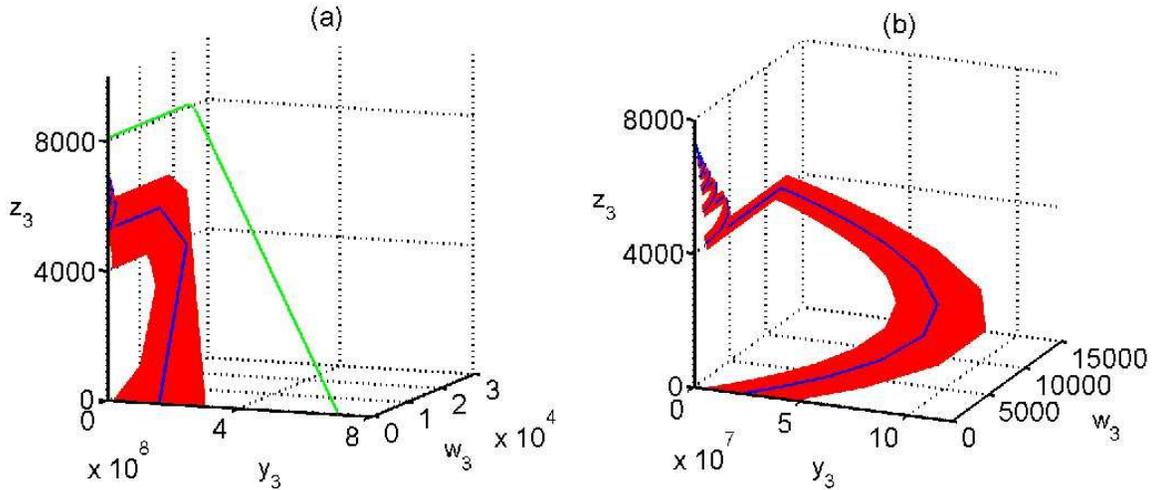,width=17cm}
\caption{(a) Reduced phase space of the system (\ref{vs}). (b) Close-up near the hypersurface of equilibria. Parameter values are the same as in Fig.~\ref{fig2}.}\label{fig3}
\end{figure}

If during time evolution, the trajectory reaches the hyperplane
$z_{i}=\phi/\alpha$ for some $i$, and at least one of $y_{i}$ or
$w_{i}$ is different from zero, then this trajectory will escape
the basin of attraction of the point $({\bf y}={\bf w}=0,{\bf
z}=(\phi/\alpha){\bf 1}_{N})$ and instead it will asymptotically
converge to the hypersurface of equilibria with the value of
$z_i>\phi/\alpha$ without any further phase space excursions. This will happen provided
the initial amount of a given variant is high enough. Figure~\ref{fig3} (a) shows in red
the projection of the stable manifold of the point on the reduced phase space of a single variant
together with a representative trajectory in blue. In the same figure a trajectory in green illustrates the scenario in which
the protective level of immune response is reached within one parasitemia peak, and hence there are no further oscillations. In Fig.~\ref{fig3} (b) we show the close-up of the phase dynamics in the neighbourhood of the hypersurface of equilibria. One can clearly observe recurrent oscillations of parasitemia, during which the specific immune response is
monotonically increasing until it reaches the protective level.

Next we would like to discuss the issues of peak dynamics and the threshold
for chronicity, which have been previously studied in Recker and Gupta
\cite{RG}. The chronicity threshold is defined as the critical ratio of the variant destruction rates $\gamma_{c}=\alpha'/\alpha$, such that if $\gamma<\gamma_{c}$, then during the first peak the protective level of immunity will be reached, so that the system will display no further oscillations.
There are several simplifying assumptions, which have to be made in
order to derive analytical expressions for the solutions needed for the analysis of peak dynamics. First of all, it is reasonable
to assume that all variants in the full system (\ref{vs}) are identical, and therefore this system can be replaced by
\[
\begin{array}{l}
\dot{y}=y(\phi-\alpha z-\alpha' w),\\
\dot{z}=\beta y,\\
\dot{w}=\beta' n_{c}y-\mu' w,
\end{array}
\]
where $n_{c}$ is the number of connections for each variant, and $y_{i}=y$, etc. The second assumption  is that for a single parasitemia peak the cross-reactive immune response does not have time to decay, i.e. for a peak dynamics we have $\mu'=0$. This reduces the system to
\begin{equation}\label{fss}
\begin{array}{l}
\dot{y}=y(\phi-\alpha z-\alpha' w),\\
\dot{z}=\beta y,\\
\dot{w}=\beta' n_{c}y.
\end{array}
\end{equation}
Assuming zero initial conditions $z(0)=w(0)=0$, which correspond to the absence of pre-existing specific or cross-reactive immune responses, the
analytic expression for the solutions of the system (\ref{fss}) can be found as
\begin{equation}\label{anal_soln}
\begin{array}{l}
\displaystyle{z(t)=\frac{1}{\psi}\left[\phi+\sqrt{2C_{1}\psi}\tanh\left(\sqrt{\frac{C_1\psi}{2}}(t+C_2)\right)\right],}\\\\
\displaystyle{y(t)=\frac{C_1}{\beta}\left[1-\tanh\left(\sqrt{\frac{C_1\psi}{2}}(t+C_2)\right)^{2}\right],}
\end{array}
\end{equation}
where the integration constants $C_1$ and $C_2$ are given by
\[
C_1=\beta y_0+\frac{\phi^2}{2\psi},\mbox{
}C_2=-\sqrt{\frac{2}{C_1\psi}}\arctanh\left(\sqrt{\frac{C_1-\beta y_0}{C_1}}\right),
\]
and $\psi$ is defined as
\[
\psi=\frac{\alpha\beta+\alpha' n\beta'}{\beta}.
\]
Initially, $y(t)$ monotonically increases, until it reaches its peak of $y=C_{1}$ exactly at $t=-C_{2}$, after which $y(t)$ is monotonically decreasing. Due to the symmetry of the solution, at $t=-2C_{2}$, $y$ has the same value as it had at the initiation of parasitemia peak. By considering the equation for $y(t)$, one can argue that if at the end of the parasitemia peak the combined specific and cross-reactive immune response has reached the protective level of $\phi/\alpha$, then this will prevent further oscillations. Evaluating $z$ and $w$ at the end of parasitemia peak, we find the threshold for chronicity as
\[
z(-2C_{2})+w(-2C_{2})>\phi/\alpha\hspace{0.5cm}\mbox{if}\hspace{0.5cm}\gamma<\gamma_{c}=
\frac{\alpha'}{\alpha}=2+\frac{\beta}{n_{c}\beta'}.
\]
If $\gamma<\gamma_{c}$, then during the first peak $y$ should be sufficiently high to allow the build-up of protective immunity. Conversely, if $\gamma>\gamma_{c}$, then $y$ will be too low for protective immunity to be reached within one peak, and therefore the system will display further oscillations \cite{RG}.

It is important to note that the trajectories shown in red and blue in Fig.~\ref{fig3} satisfy the condition for chronicity $\gamma<\gamma_{c}$,  but still for these trajectories the protective level of immune response is not reached within a single parasitemia peak. The reason for this discrepancy is due to the fact that in the system describing the peak dynamics, the cross-reactive immune response does not decay, because if it did, then at the end of the parasitemia peak the combined immunity would be below the protective level, and therefore further oscillations would occur, as shown in Fig.~\ref{fig3}. This also highlights the importance of initial conditions, and in particular, the initial amount of variants, which may play a crucial role in whether or not the protective immunity level will be reached within
one peak for the same parameter values.

\section{General case}

In the previous section we considered the case $\mu=0$, in which
the long-lasting immune responses can only grow with time, unless
they are saturated at the level preventing further re-emergence of
particular variants. For $\mu>0$, the situation is drastically
different as the hypersurface of equilibria no longer exists, and
instead, it degenerates into two separate steady states. One of
these is the origin $\left({\bf y}={\bf z}={\bf w}=0\right)$,
which is always a saddle with an $N$-dimensional unstable manifold
and a $2N$-dimensional stable manifold. The other steady state
originating from the hypersurface of equilibria is the fully
symmetric equilibrium
\begin{equation}
\begin{array}{l}
\displaystyle{y_i=\frac{\phi\mu\mu'}{\alpha\beta\mu'+\alpha'n_{c}\beta'\mu},\mbox{
}z_i=\frac{\phi\beta\mu'}{\alpha\beta\mu'+\alpha'n_{c}\beta'\mu},\mbox{
}}\\\\
\displaystyle{w_i=\frac{\phi\mu n_{c}
\beta'}{\alpha\beta\mu'+\alpha'n_{c}\beta'\mu},\mbox{
}i=\overline{1,N},}
\end{array}
\end{equation}
where $n_{c}=m+n-1$ is the number of connections for each variant.
Using Fig. 1, this number can easily be interpreted as the number
of elements in the horizontal and vertical strata, to which the
current variant belongs. When considered in the context of a
reduced system (\ref{fss}), in which all variants are assumed to
behave in the same manner (see next section for further analysis
of this case), this steady state is stable for all values of
parameters, as shown in Recker and Gupta \cite{RG}. At the same
time, this result does not hold for the full system (\ref{vs}), as the stability of the fully symmetric equilibrium
does depend on parameters of the system. More specifically, the fully symmetric equilibrium can undergo
Hopf bifurcation, thus giving rise to periodic occurrences of parasitemia peaks.

\begin{figure}
\hspace{-0.5cm}\epsfig{file=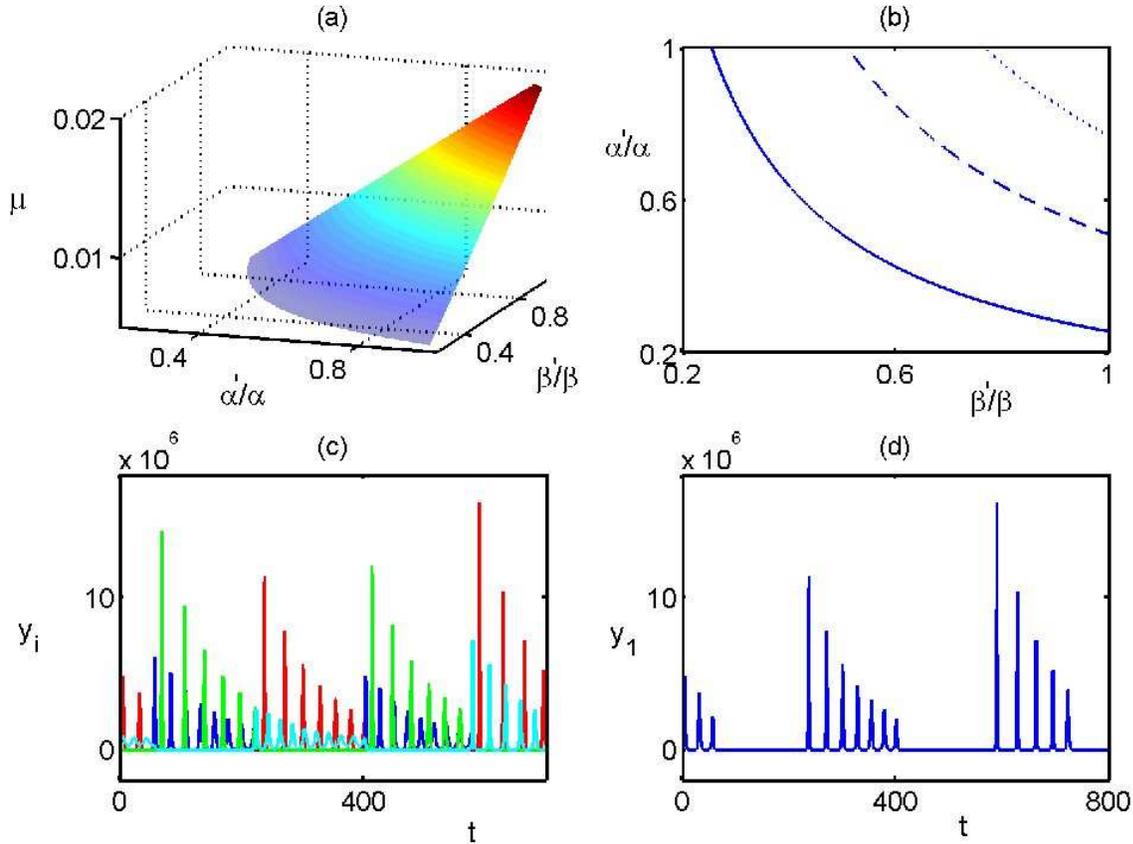,width=17cm}
\caption{a) Boundary of Hopf
bifurcation in a parameter space. b) The two-parameter Hopf
boundaries for different values of the specific response decay
rate $\mu=0.005$ (solid), $\mu=0.01$ (dashed) and $\mu=0.015$
(dotted). In (a) the fully symmetric steady state is
unstable above the coloured surface, and in (b) it is unstable above each of the corresponding curves.
(c) Temporal dynamics of variants 1 (red), 2 (blue), 3 (green) and 4 (cyan). (d) Temporal dynamics of variant 1 showing periodic oscillations with large intermittent period of quiescence.}\label{fig4}
\end{figure}

Figures~\ref{fig4} (a) and (b) show the boundary of the Hopf bifurcation in the
parameter space of the system (\ref{vs}) with two variants in each
of the two minor epitopes. These figures indicate that the higher
is the decay rate of variant specific immune response, the larger
should be the values of the relative immune efficiency
$\gamma=\alpha'/\alpha$ and that of a ratio of proliferation rates
$\beta'/\beta$ to guarantee the occurrence of the Hopf
bifurcation. The corresponding temporal evolution of variants in the parameter regime beyond the Hopf bifurcation
is illustrated in figures (c) and (d). One can observe periodic oscillations of all four variants, which have approximately
the same maximum amplitudes and are slightly out-of-phase with each other. 
The plot of the dynamics of one variant shown in Fig.~\ref{fig4} (d) indicates that peaks of parasitemia corresponding to this variant have decreasing amplitudes, and after several occurrences there are large periods of time when the variant is quiescent. Cross-reactivity between different variants causes subsequent re-appearance of large-amplitude oscillations after such periods of quiescence. The reason for this is as follows. The long-lasting and cross-reactive immune responses show anti-phase oscillations, which are quite regular both in amplitude and in period. These oscillations lead to a slightly irregular oscillations of the combined rate of variant destruction $\alpha z+\alpha' w$. Intervals of parasitemia peaks correspond to the combined variant destruction rate oscillating around the critical value of $\phi$ with long-lasting immune response increasing. After such intervals, the combined variant destruction rate stays above $\phi$ keeping the variant absent from the dynamics, and during this time the long-lasting immune response wanes, until it starts to recover during the next cycle. We emphasize that this dynamics can only occur in the case when the long-lasting immune response can decay, hence this feature could not be observed in the previously analysed case of $\mu=0$.

Besides the origin and a fully symmetric equilibrium, the system
also possesses $(2^{N}-2)$ steady states characterized by a
different number of non-zero variants $y_i$. One should notice that
the symmetry of the system mentioned earlier implies that for a
given number of non-zero variants, many of the corresponding steady
states are symmetry-related. At the same time, one can identify
several clusters of the steady states with different values of the
steady states which cannot be transformed into each other by a
symmetry. For example, if we consider the system with two variants in each of the two minor epitopes,
then there exist six steady states with two non-zero variants. Introducing the notation
\[
Y_{1}=\frac{\phi\mu\mu'}{\beta'\alpha'\mu+\alpha\beta\mu'},
\hspace{0.5cm}Y_{2}=\frac{\phi\mu\mu'}{2\beta'\alpha'\mu+\alpha\beta\mu'},
\]
the steady states with non-zero variants 12, 13, 24 and 34 form one cluster:
\[
\begin{array}{l}
E_{12}=(Y_{2},Y_{2},0,0,Z_{2},Z_{2},0,0,2W_{2},2W_{2},W_{2},W_{2}),\\
E_{13}=(Y_{2},0,Y_{2},0,Z_{2},0,Z_{2},0,2W_{2},W_{2},2W_{2},W_{2}),\\
E_{24}=(0,Y_{2},0,Y_{2},0,Z_{2},0,Z_{2},W_{2},2W_{2},W_{2},2W_{2}),\\
E_{24}=(0,0,Y_{2},Y_{2},0,0,Z_{2},Z_{2},W_{2},W_{2},2W_{2},2W_{2}),
\end{array}
\]
while the steady states with non-zero variants 14 and 23 are in another cluster
\[
\begin{array}{l}
E_{14}=(Y_{1},0,0,Y_{1},Z_{1},0,0,Z_{1},W_{1},2W_{1},2W_{1},W_{1}),\\
E_{23}=(0,Y_{1},Y_{1},0,0,Z_{1},Z_{1},0,2W_{1},W_{1},W_{1},2W_{1}),
\end{array}
\]
where $Z_{1,2}=\beta Y_{1,2}/\mu$ and $W_{1,2}=\beta' Y_{1,2}/\mu'$. All the steady states in the first cluster are related by permutation, and the steady states in the second cluster are also related by some permutation, but the steady states from the first cluster cannot be related to those in the second cluster. The reason for this becomes clear if one more closely analyses the structure of the adjacency matrix $A$ given in (\ref{mat4}). In the case of a steady state $E_{ij}$ from the first cluster, both rows $i$ and $j$ of matrix $A$ contain ones in positions $i$ and $j$ (i.e. the variants $i$ and $j$ cross-react with each other), while in the case of the second cluster the rows $i$ and $j$ contain only a single one in either position $i$ or position $j$ (i.e. the variants $i$ and $j$ are completely unrelated). Due to this difference the steady states from the two clusters are different and it is impossible to change from one cluster to another by permutation.

As far as stability of the steady
states different from the origin and the fully symmetric equilibrium is concerned, they all
are saddles of different dimensions. Even though they are unstable
as steady states, it is possible for some of them to form some sort of a heteroclinic cycle.\\

\noindent {\bf Remark.} In the case when the number $N$ of malaria
variants participating in the dynamics exceeds four, the symmetry
of the system increases the co-dimension of the Hopf bifurcation
for the fully symmetric steady state. Moreover, the purely
imaginary eigenvalues at the Hopf bifurcation would coincide, thus
creating extra complications for the analysis by virtue of
increasing the dimension of the centre manifold. Some details of
possible bifurcation scenarios in systems with an "all-to-all"
coupling can be found in \cite{DS,E}, and the extension of those
results should provide an insight into the effects of symmetry on
the dynamics of system (\ref{vs}). The complete analysis of these
effects will be presented elsewhere.

\section{Robustness of the fully symmetric solution}

An interesting dynamical regime occurs when, by virtue of initial
conditions or time evolution, the system behaves in such a way
that all variants are indistinguishable from each other, in other
words, the system is in a state of complete symmetry. In this
case, the dimension of the system reduces drastically from $3N$ to
just three. As several insightful results have been obtained for
this case \cite{RG}, it is important to study how robust this
state of complete symmetry is with respect to perturbations that
attempt to break the symmetry. To characterize stability
properties of the symmetric state one can use transverse Lyapunov
exponents, as is customary in the studies of synchronization, see,
for instance \cite{PC}. By analogy with synchronization theory we
shall call the hypersurface of complete symmetry a {\it symmetry
manifold}.

Writing
$(y_i=y+\widetilde{y}_i,z_i=z+\widetilde{z}_i,w_i=w+\widetilde{w}_i)$,
one can split the total dynamics into that inside the symmetry
manifold
\begin{equation}\label{reduced}
\begin{array}{l}
\dot{y}=y(\phi-\alpha z-\alpha' w),\\
\dot{z}=\beta y-\mu z,\\
\dot{w}=\beta' n_{c}y-\mu' w,
\end{array}
\end{equation}
and the linearized dynamics in the transverse direction given by
\begin{equation}\label{lin_dyn}
\begin{array}{l}
\displaystyle{\frac{d}{dt} \left(
\begin{array}{l}
\widetilde{{\bf y}}\\
\widetilde{{\bf z}}\\
\widetilde{{\bf w}}
\end{array}
\right)=DF(y{\bf 1}_{N},z{\bf 1}_{N},w{\bf 1}_{N})\left(
\begin{array}{l}
\widetilde{{\bf y}}\\
\widetilde{{\bf z}}\\
\widetilde{{\bf w}}
\end{array}
\right),}\\\\
DF(y{\bf 1}_{N},z{\bf 1}_{N},w{\bf 1}_{N})=\left(
\begin{array}{ccc}
(\phi-\alpha z-\alpha' w){\bf I}_{N}&-\alpha y{\bf I}_{N}&-\alpha'
y{\bf I}_{N}\\
\beta{\bf I}_{N}&-\mu{\bf I}_{N}&{\bf 0}_{N}\\
\beta' A&{\bf 0}_{N}&-\mu'{\bf I}_{N}
\end{array}
\right).
\end{array}
\end{equation}
Here $n_{c}$ is again the number of connections of a given
variant, ${\bf 0}_{N}$ and ${\bf I}_N$ denote $N\times N$ zero and
unity matrices, respectively. The minimal condition for the
stability (or robustness) of the symmetric state is that the
maximum Lyapunov exponent associated with the system
(\ref{lin_dyn}) has negative real part \cite{PC}. By solving
equations (\ref{lin_dyn}) in combination with (\ref{reduced}), we
determine the dependence of the leading Lyapunov exponent on the
system parameters.

\begin{figure}
\hspace{-1cm}\epsfig{file=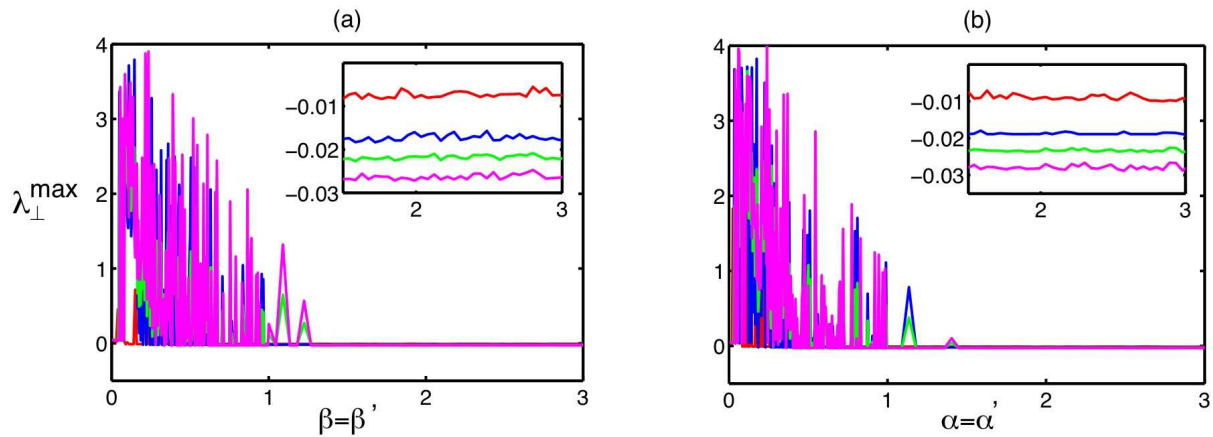,width=17cm}
\caption{Maximal
transverse Lyapunov exponent as a function of parameters. a)
Parameter values are $\phi=7.5$, $\beta=\beta'=1$,
$\mu'=0.02$. b) Parameter values are $\phi=7.5$,
$\alpha=\alpha'=1$, $\mu'=0.02$. Different colours represent
different decay rates of the long-lasting immune response: $\mu=0$
(red), $\mu=0.01$ (blue), $\mu=0.02$ (green) and $\mu=0.03$
(magenta).}\label{fig5}
\end{figure}

In Figure~\ref{fig5} we show the results of numerical simulations
for the maximal transverse Lyapunov exponent $\lambda_{\bot}^{\rm
max}$. In both plots we kept the rates of variant destruction equal to each other $\alpha=\alpha'$ and also
the proliferation rates were taken to be the same $\beta=\beta'$. Figure ~\ref{fig5} indicates that for small values of $\alpha=\alpha'$ or $\beta=\beta'$, the fully symmetric state of the system is transversely unstable, as signified by
the positive transverse Lyapunov exponent. This means that in such parameter regime different variants will not
synchronize in time, hence it is unlikely to observe the fully symmetric state in experiment. However, when the variant destruction rates/proliferation rates are increased, the fully symmetric state becomes transversely stable, i.e. independently on initial conditions for each particular variants, they will all ultimately follow the same time evolution. The increase in the decay rate of the specific immune response $\mu$ plays a stabilizing role, since it lowers the values of of the maximal transverse Lyapunov exponent. The robustness of the fully symmetric state appears to be independent on the relative efficiency $\gamma=\alpha'/\alpha$ of immune responses.

\section{Conclusions}

In this paper the temporal behaviour in a model of antigenic
variation in malaria has been studied from a dynamical systems
perspective. Using the model of immune response to multiple
epitopes, we have demonstrated that when the long-lasting immune
response does not decay, the system possesses a high-dimensional
surface of equilibria, and these exhibit steady-state bifurcation
without parameters, i.e. some part of the surface of equilibria
consists of saddles of different dimensions, while another part
contains stable nodes. The existence of these two parts of the
surface of equilibria with different stability properties accounts
for the observed patterns of behaviour of malaria variants, when
different variants exhibit out-of-phase parasitemia peaks that decay
with time. If the initial amounts of all variant are not very large, then phase space excursions between successive
re-appearances of the variants become shorter as the time grows, and
eventually all trajectories approach the single steady state {\bf T}
characterized by all coordinates equal to each other and equal to
the value at the boundary between the saddles and the nodes on the
surface of equilibria. If, however, an initial amount of a given variant is sufficiently high, then a trajectory
with such initial condition will escape the basin of
attraction of the above-mentioned point {\bf T} by reaching the protective level of
long-lasting immune response to a given variant while having either
a non-zero transient response to this variant or a non-zero amount of
the variant itself. In this case, the eventual time evolution of the solution will be different in that
it will also approach the surface of equilibria but now it
will be above the critical protective level without any further
excursions in the phase space.

When both variant-specific and cross-reactive immune responses are
allowed to decay with a certain rate, the dynamics are quite
different. In this case the surface of equilibria disintegrates,
and instead the phase space of the system contains a large number
of distinct fixed points many of which are related to each other
by the permutation symmetry of the variants. At the same time,
they may form separate clusters which are not related by symmetry. Provided the decay rate of the specific immune response is high enough,the fully symmetric equilibrium
will exhibit Hopf bifurcation, thus giving rise to periodic oscillations of the variants. These oscillations
appear to be out-of-phase for different variants, and such oscillations are separated by extended
time intervals during which the amount of a variant is very small.

In order to investigate to what extent the results obtained in the
approximation of complete symmetry between variants describe the
general patterns of behaviour, we have numerically computed the
transverse Lyapunov exponents of the fully symmetric state. This
analysis indicates that while the fully symmetric state
 is not robust to small perturbations for small proliferation/variant destruction rates,
the robustness is restored as these rates increase. In this case the dynamics of the completely symmetric system
faithfully represents that of the full original system. Finally, we note that
the robustness of complete synchronization between variants
increases with the decay rate of the specific immune response.

\section*{Acknowledgements} The authors would like to thank Marty Golubitsky,
Hinke Osinga, Oleksandr Popovych and Mario Recker for useful
discussions. They would also like to thank two referees for their comments and suggestions, which have helped
to improve the presentation in this paper.


\begin{thebibliography}{50}

\bibitem{AF} Ashwin, P., Field, M.: Heteroclinic networks in coupled
cell systems. {\it Arch. Rat. Mech. Anal.} {\bf 148}, 107-143
(1999).

\bibitem{DG} Dawes, J.H.P., Gog, J.R.: The onset of oscillatory
dynamics in models of multiple disease strains. {\it J. Math.
Biol.} {\bf 45}, 471-510 (2002).

\bibitem{DS} Dias, A.P.S., Stewart, I.: Secondary nifurcations in
systems with all-to-all coupling. {\it Proc. R. Soc. Lond.} A {\bf
459}, 1969-1986 (2003).

\bibitem{E} Elmhirst, T.: ${\bf S}_{N}$-equivariant
symmetry-breaking bifurcations. {\it Int. J. Bif. Chaos} {\bf 14},
1017-1036 (2004).

\bibitem{FLA} Fiedler, B., Liebscher, S., Alexander, J.C.: Generic
Hopf bifurcation from lines of equilibria without parameters. {\it
J. Diff. Eqns.} {\bf 167}, 16-35 (2000).

\bibitem{GS} Golubitsky, M., Schaeffer, D.: Singularities and
groups in bifurcation theory. Springer-Verlag, New York (1985).

\bibitem{Gu} Gupta, S.: Parasite immune escape: new views into
host-parasite interactions. {\it Curr. Opin. Microbiol.} {\bf 8},
428-433 (2005).

\bibitem{PC} Pecora, L.M., Carroll, T.L.: Synchronization in chaotic
systems. {\it Phys. Rev. Lett.} {\bf 64}, 821-824 (1990).

\bibitem{PD} Postlethwaite, C., Dawes, J.H.P.: Regular and irregular
cycling near a heteroclinic network. {\it Nonlinearity} {\bf 18},
1477-1509 (2005).

\bibitem{RNBKMNG} Recker, M., New, S., Bull, P.C., Linyanjui, S.,
Marsh, K., Newbold, C., Gupta, S.: Transient cross-reactive immune
responses can orchestrate antigenic variation in malaria. {\it
Nature} {\bf 429}, 555-558 (2004).

\bibitem{RG} Recker, M., Gupta, S.: Conflicting immune responses can
prolong the length of infection in {\it Plasmodium falciparum}
malaria. {\it Bull. Math. Biol.} {\bf 68}, 821-835 (2006).

\bibitem{RTO} Rokni Lamooki, G.R., Townley, S., Osinga, H.M.:
Bifurcations and limit dynamics in adaptive control systems. {\it
Int. J. Bif. Chaos} {\bf 15}, 1641-1664 (2005).

\bibitem{SR} Shampine, L.F., Reichelt, M.W.: The MATLAB ODE Suite. {\it SIAM J. Sci. Comp.} {\bf 18}, 1-22 (1997).

\end{thebibliography}
\end{document}